\documentclass{article}

\usepackage{spconf}
\usepackage{graphicx}
\usepackage{amsmath, amsthm, amssymb, amsfonts}
\usepackage{algorithm, algorithmic, ifsym, subfigure, bm}
\usepackage{mathtools}
\usepackage{enumitem}
\usepackage{epstopdf}

\def\bc{ \textbf{c} } 
\def\bk{ \textbf{k} }

\def\bA{ {\mathbf{A}} }
\def\bB{ {\mathbf{B}} }

\def\bK{ {\mathbf{K}} }

\def\bP{ {\mathbf{P}} }
\def\bQ{ {\mathbf{Q}} }
\def\bR{ {\mathbf{R}} }

\def\calA{{ \mathcal{A}  }}
\def\calS{ \mathcal{S} }

\def\Tdim{M}
\def\Sca{J}
\def\auxdim{V}

\def\Tslot{ T_{slot} }
\def\Tps{T_{psense}}

\def\balpha{ {\bm{\alpha}} }

\def\nn{{ \parallel   }}

\def\RRnn{{ \mathbb{R}_+  }}

\def\EE{{ \mathbb{E}  }}

\def\NNz{{ \mathbb{N}_0 }}

\def\ba{ \textbf{a} } 
\def\bat{ \tilde{\textbf{a}} } 
\def\bs{{ \mathbf{s}  }}
\def\bst{ \tilde{\textbf{s}} } 

\def\bx{{ \mathbf{x}  }}
\def\bxt{{ \tilde{\mathbf{x}}  }}

\usepackage{cite} % Allows a range of citations to be expressed as, e.g. [2]-[5], instead of [2],[3],[4],[5]

\newcommand{\defequal}{ \stackrel{\rm def}{=}  }

\usepackage{subfig}

\usepackage{float}% http://ctan.org/pkg/float
\title{Reinforcement Learning with Budget-Constrained Nonparametric Function Approximation for Opportunistic Spectrum Access}
\name{Theodoros Tsiligkaridis, David Romero \thanks{DISTRIBUTION STATEMENT A. Approved for public release. Distribution is unlimited. This material is based upon work supported under Air Force Contract No. FA8702-15-D-0001. Any opinions, findings, conclusions or recommendations expressed in this material are those of the author(s) and do not necessarily reflect the views of the U.S. Air Force.}}
\address{MIT Lincoln Laboratory}
\begin{document}
\ninept
\maketitle
\begin{abstract}
Opportunistic spectrum access is one of the emerging techniques for maximizing throughput in congested bands and is enabled by predicting idle slots in spectrum. We propose a kernel-based reinforcement learning approach coupled with a novel budget-constrained sparsification technique that efficiently captures the environment to find the best channel access actions. This approach allows learning and planning over the intrinsic state-action space and extends well to large state spaces. We apply our methods to evaluate coexistence of a reinforcement learning-based radio with a multi-channel adversarial radio and a single-channel CSMA-CA radio. Numerical experiments show the performance gains over carrier-sense systems.
\end{abstract}

\section{Introduction} \label{sec:intro}
It has been widely recognized that static spectrum allocation methods lead to spectrum being under-utilized \cite{FCC:2002}. To help remedy this issue as demand for spectrum resources continues to increase, opportunistic spectrum access (OSA) can be used to efficiently exploit spectrum opportunities by allowing secondary users (SU) within a \textit{cognitive radio network} (CRN) to find and access under-utilized parts of the spectrum and cause minimal interference to primary users (PU) \cite{Zhao:2007}. 

Much of the current literature on OSA deals with its underlying CRN architecture and implementation \cite{Sutton:2010, Xu:2012, Soltani:2015, Wang:2016, Rawat:2016}, performance optimization and MAC protocol design \cite{Zhao:2007:b, Chen:2008, Zhao:2008:a, Zhao:2008:b, Derakhshani:2012, Jayaweera:2015}, but less effort has been given to developing \textit{learning}-based approaches that are effective and practical for use in unknown RF environments in which the PU reacts to SU transmissions. Hidden Markov models, neural networks, and other predictive models tend to require a prohibitive amount of computational resources to train, often require periodic retraining, and may require a very large number of parameters to capture relevant communication environments \cite{Browne:2012, Tumuluru:2012}. Furthermore, deep learning models are sensitive to noise \cite{Nguyen:2015} and adversarial actions \cite{Papernot:2016}. These problems prohibit the use of these algorithms in practical environments due to the amount of data and tight latency requirements. For spectrum prediction and exploitation, this data must be analyzed in real-time to be valuable for decision making.

In this paper, we use reinforcement learning (RL) \cite{SuttonBarto} to enable OSA in unknown and dynamic RF environments. In our problem formulation, SU channel accesses change the RF environment due to PU coexistence protocols, e.g., collisions may force a PU to change frequencies or backoff their data rate. RL provides a natural framework for model-free optimization in which radios learn by direct interaction with the environment. RL has been used in  cognitive radio networks in the context of cooperative sensing, interference control, optimization of queueing systems, and spectrum management \cite{Lo:2013,Galindo:2010,Wang:2014,Morozs:2016}.

%The exploration-exploitation tradeoff is a well known issue that must be addressed when considering the use of RL\cite{SuttonBarto}.
%To learn a good spectrum access policy, radios must balance the exploration of the space of states and actions with the exploitation of an already learned policy \cite{SuttonBarto}. As radios use acquired knowledge to choose actions that maximize a cumulative reward, a dictionary of state-action pairs is built in a nonparametric fashion. 
When applying RL to OSA, widely used methods for balancing the exploration-exploitation tradeoff \cite{SuttonBarto} can lead to slow learning. Furthermore, in time varying RF environments, the space of states and actions can be very large, which can lead to poor generalization, and further slowdown the learning process. We remedy these problems by developing a heuristic for efficient exploration, and a novel budget-constrained dimensionality reduction method. Our approach allows for quick adaptation, high accuracy and manageable computational complexity. 

The contributions of this paper are as follows. 
\begin{itemize}
\item We provide a heuristic method for guided exploration of the state-action space that accelerates learning. 
\item We propose a new criterion based on memory and kernel principal angles for removal of a dictionary item that couples well with the kernel sparsification criterion. 
\item We apply our kernel-based RL techniques on a radio that coexists with an adversarial multi-channel radio and a single-channel CSMA-CA radio and observe significant gains over carrier-sense systems.
\end{itemize}

%The algorithms developed in this work may also be useful for prediction in high-dimensional data, adaptive sensing and stochastic control problems.

\section{Related Work}\label{sec:relwork}
%RL for OSA has received much attention in the literature,\cite{OKSANEN2012}, \cite{LundenJSTSP2012}, \cite{Alsaleh2011}, \cite{Wu2010}, \cite{Lo:2013}, and \cite{LundenSPM2015}. There are a variety of ways in which the problem has been formulated and approached. For example, OSA is cast as sensing policy optimization in \cite{OKSANEN2012,Lo:2013}, where the authors propose cooperative Q-learning based approaches. Single and multi-agent Q-learning for OSA is considered in \cite{Alsaleh2011}. Multi-agent RL for OSA is considered in \cite{LundenJSTSP2012} and \cite{Wu2010}, where function approximation is used to handle high dimensional state-action spaces. The authors of \cite{Morozs:2016} propose a heuristically accelerated Q-learning approach to spectrum management in LTE cellular systems. 

Our approach to RL for OSA uses Q-learning with budget constrained function approximation and guided exploration for learning acceleration. Many previous works have considered Q-learning for OSA, e.g. \cite{Alsaleh2011,Morozs:2016,Wu2010}, and \cite{Lo:2013}. The majority of existing approaches consider finite state and action spaces where the Q-function is stored in a table \cite{Alsaleh2011,Morozs:2016,Lo:2013}, whereas we consider continuous state spaces that are handled using kernel based function approximation. RL with function approximation is considered in \cite{LundenJSTSP2012,Wu2010}. However, the authors of \cite{Wu2010} consider a different function approximation approach, and their state space is defined differently from ours. Also, the state transition function considered in \cite{LundenJSTSP2012} must be either known or estimated and does not depend on SU actions. In contrast, we consider state transition functions that react to SU actions, are unknown and need not be directly estimated. The guided exploration approach we propose for learning acceleration is related to \cite{Morozs:2016}, but the authors of \cite{Morozs:2016} use a stateless form of Q-learning, and the heuristic function proposed for accelerated learning is less general than the one proposed in this work. The kernel sparsification method we employ for adding elements to the state action dictionary was considered by \cite{Xu:2007} within the context of policy iteration based RL using batch processing. In contrast, we consider an online approach and propose a method for removing elements from the state action dictionary that must remain bounded in size.

\section{Problem Formulation} \label{sec:model}
The objective of the SU radio network is to find a channel access policy that maximizes its throughput while avoiding collisions with a PU network. A Markov decision process (MDP) $(\calS,\calA,R,\bP)$ may be used to represent the dynamic communication environment of the PU network, where $\calS$ is the state space, $\calA$ is the action space, $R$ is the reward function and $\bP$ encodes the state-transition dynamics. A \textit{policy} is a function $\pi: \calS \to \calA$ that maps states to actions. For simplicity, we consider a single SU receiver that may be used to model a single link with another SU radio, or a centralized controller that learns and distributes a channel access schedule to user pairs via a control channel, assuming all SU radios are experiencing similar path losses.

\subsection{State Space/Observation Model}
The SU receiver observes the state of the RF environment as a sequence of discrete time steps $t = 0,1,2,\cdots$, where the observed state at time $t$ is denoted $\bs_t$. To construct each state, we assume the received bandwidth is divided into $K$ non-overlapping $W$ Hz channels, and the considered temporal window is segmented into $\Tdim$ non-overlapping time slots of duration $\Tslot$. Furthermore there may be up to $\auxdim$ auxiliary terms included in each state. The receiver senses the entire band and updates any auxiliary terms, yielding a state $\bs_t \in \calS$ where $\calS \defequal \RRnn^{K\Tdim+\auxdim}$. 

% The states are power measurements of noise plus any signals received from the environment. 
\subsubsection{Sensing and Slot Structure}\label{sec:senseslots}
\begin{figure}%[ht]
	\centering
		\includegraphics[width=0.40\textwidth]{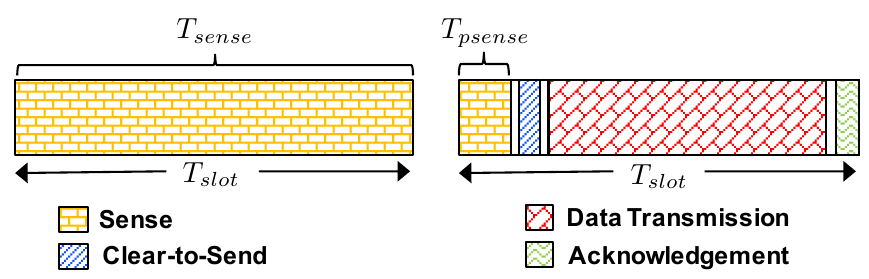}
	\caption{ \small{Half-duplex slot structure for each frequency channel. Left (no SU access): entire slot duration is sensed. Right (SU access): slot is partially sensed.} }
	\label{fig:hd_slot1}
\end{figure}

We assume the SU radios operate in half-duplex mode. During each time slot, all $K$ channels are either partially or persistently sensed and the average observed power is computed for each channel. Partial slot sensing occurs for the channel-slots that the SU attempts to access, where only a fraction of the entire slot, $\Tps <  \Tslot$, is sensed (see Fig. \ref{fig:hd_slot1}). Note that $\bs_t$ is updated at $\Sca\Tslot$ spaced intervals, where $\Sca$ is the temporal dimension of the action space.% to be described next.%States $s_t$ are constructed as vectors of average power measurements are constructed using a finite length sliding window of successive measurements.

\subsection{Action Space}
The action space $\calA$ is finite and can be pre-specified or learned in a nonparametric manner. For the multichannel case, $\calA \subset \{0,1\}^{KJ}$, the RL radio actions are chosen to be relevant for the environment the radio is operating in. This is accomplished by interacting with the PU network for a short time and building up the action space by including actions as the inverse of the detected PU communication patterns, for which $|\calA| \ll 2^{KJ}$.

For the single channel case, each time an action is to be selected the SU either accesses the channel or remains idle for the following $\Sca$ consecutive time slots, yielding $|\calA|=2$. A channel access at time $t$ using transmit power $P_{stx}$ is denoted $\ba_t = P_{stx}\mathbf{1}_\Sca$, whereas remaining idle is denoted $\ba_t = \mathbf{0}_\Sca$. We remark that in the single channel case, larger action spaces that enable a more rich set of channel access strategies are possible within our learning framework. However, such action spaces generally tend to incur longer training times, and are differed to future extensions of the current work.

%may access one of the $K$ channels at each time, and makes decisions $a_t \in \calA$ about which channel to access based on the current state $s_t$. The action being taken $a_t$ affects in general the evolution of the states. Of course, this assumption can be relaxed and the methods presented in the paper are still valid as long as the action space is parameterized appropriately. In the most general case where the blue link is capable of accessing up to $K_a$ out of $K$ channels at each time, action can be represented as binary vectors in a space $\calA$ of dimension $|\calA|=\sum_{k=1}^{K_a} {K \choose k}$.

\subsection{Transition, Reward and Noise Models}
The underlying system (i.e., described by the PU communication patterns) evolves from a state $\bs_t$ to a state $\bs_{t+1}$ probabilistically when the SU radio takes an action $\ba_t$. This transition model does not need to be known or explicitly learned by the SU network radios.

%\subsection{Reward Model}
The reward obtained when transitioning to state $\bs_{t+1}$ as a result of taking action $\ba_t$ in state $\bs_t$ is proportional to the number of bits received successfully and in practice can be measured by an acknowledgment. If the packet fails to be successfully received, a negative reward is incurred due to the resulting collision with the PU network.
%This can be written as:
%\begin{equation*}
	%r_t = \begin{cases} R({\ba_t}), & \text{if channels assoc. w. action $\ba_t$ are idle at state $\bs_{t+1}$}  \\ -C_s, & \text{else} \end{cases}
%\end{equation*}
%where $R(a)$ is the number of bits successfully received when using channels $a$. %Note that we make the assumption that all packet losses for the blue network are due to channel access collisions.
%\textcolor{red}{This section needs attention to ensure the description of rewards is compatible with both multi-channel and single channel cases.}

%\subsection{Noise Modeling} \label{sec:noise}
We assume an additive white Gaussian noise (AWGN) channel model at the SU. Based on the noise level in the channel, we use a worst case\footnote{Note that $P_d$ and $P_f$ vary depending on whether a given slot is persistently or partially sensed. } probability of detection $P_d=0.996$ and false alarm $P_f=0.001$.

%In practice, there will be noise due to the receiver which makes the true states unobservable. The observation model in a AWGN channel over a subband $k$ at time $t$ is:
%\begin{equation*}
	%y_t(k) = z_t(k) + w_t(k)
%\end{equation*}
%where $w_t(k)$ is complex-valued white Gaussian noise with variance $\EE|w_t(i)|^2 = \sigma_n^2$. The true power level is $|z_t(k)|^2$ and the noiseless state-vector is $\bp_t=[|z_t(1)|^2,\dots,|z_t(K)|^2]^T$. Taking the power measurement of $y_t(k)$ and assuming independence between the transmitted signal and noise, we obtain:
%\begin{equation*}
	%\bs_t = \bp_t + \bn_t
%\end{equation*}
%where $\bn_t = [|w_t(1)|^2,\dots,|w_t(K)|^2]^T$. The noise distribution is exponential for each subband $k$, and due to independence across subbands, the multivariate distribution factors as
%\begin{equation*}
	%f(\bn) = \prod_{k=1}^K \frac{1}{\sigma_n^2} e^{-n(k)/\sigma_n^2} I_{\{n(k)\geq 0\}}
%\end{equation*}
%The noise in our power measurements $\bs_t$ is highly non-Gaussian and concentrated on the positive orthant of the $K$-dimensional channel space.

\subsection{Reinforcement Learning with Nonparametric Function Approximation}
The objective is to find the policy $\pi$ that maximizes the discounted cumulative reward $\EE^\pi \left[ \sum_{t=0}^\infty \gamma^t r_t \right], \gamma\in[0,1)$. The optimal policy is obtained as $\pi^*(\bs) = \arg \max_{\ba\in \calA} Q^*(\bs,\ba)$ where $Q^*(\bs,\ba) = \max_\pi Q^\pi(\bs,\ba)$, and $Q^\pi(\bs,\ba)$ is the value of taking action $\ba_0 = \ba$ in state $\bs_0 = \bs$ and following policy $\pi$ \cite{SuttonBarto}. 

In practice, $\calS$ may contain a prohibitively large number of states. To handle this and help with generalization, we employ nonparametric function approximation. Using the kernel-based approach of \cite{SuttonBarto, Xu:2007}, we model the state-action value functions using the nonparametric linear approximation:
\begin{equation} \label{eq:kernelQ}
	Q(\bs,\ba) = \sum_{l=1}^L \alpha_l k((\bs,\ba),(\bst_l,\bat_l))
\end{equation}
where $k(\cdot,\cdot)$ is a kernel function, $\alpha_l$ is the weight associated with state action pair $(\bst_l,\bat_l) \in \mathcal{D}$, $L = |\mathcal{D}|$, and $\mathcal{D}$ is a dictionary of state action pairs that must be learned. In the sequel, $\bst_l$ and $\bat_l$ are used to denote state action pairs contained in $\mathcal{D}$, whereas $\bs_t$ and $\ba_t$ denote observed states. In this paper we use the exponential kernel, $k((\bs,\ba),(\bst,\bat)) = \exp\left(\frac{-\nn \bs - \bst\nn^2}{2 \sigma_s^2}\right) \cdot \exp\left(\frac{-\nn \ba - \bat\nn^2}{2 \sigma_a^2}\right)$. Kernel methods transform state-action pairs, $(\bs_l,\ba_l)$, into a high-dimensional feature space with feature vectors $\phi(\bs_l,\ba_l)$. Inner products in the feature space may be computed with the kernel trick, $k(\bx,\bx') = \left<\phi(\bx), \phi(\bx')\right>$. See \cite{Xu:2007} Section III.A for more detailed information regarding the use of kernel methods.

\section{Kernel Q-Learning with Guided Exploration}\label{sec:KQL_GE}
The Q-learning algorithm \cite{SuttonBarto} seeks to find the policy $\pi(\cdot)$ that maximizes the discounted cumulative reward.
%\begin{equation*}
%	\EE^\pi \left[ \sum_{t=0}^\infty \gamma^t r_t \right]
%\end{equation*}
%where $\gamma \in [0,1)$ is the discount parameter. 
%This criterion models rewards received sooner as being more important than rewards received later. 
In the kernel setting, the algorithm estimates the optimal state-action value function $Q^*$ using a stochastic approximation:
\begin{align*}
	\balpha_{t+1} &= \balpha_t + \eta \Big(r_t + \gamma \max_{\ba'} \left\{ \balpha_t^T \bk_t(\bs_{t+1},\ba') \right\}\label{eq:alpha_update} \\
		&\qquad - \balpha_t^T \bk_t(\bs_t,\ba_t) \Big) \bk_t(\bs_t,\ba_t)
\end{align*}
where $\bk_{t}(\bx_t) = [k(\bxt_1,\bx_t),\dots,k(\bxt_{L_{t}},\bx_t)]^T$ is the kernel vector for $\bx_t = (\bs_t,\ba_t)$.

Action selection is based on a modification of the $\epsilon$-greedy policy that encourages exploration of actions that have remained unexplored:
\begin{equation} \label{eq:akql}
	\ba_t = \begin{cases} \arg \max_{a} Q_t(\bs_t,\ba) + \frac{\max_{\ba} Q_t(\bs_t,\ba)}{N_t(\bs_t,\ba)}, & \text{w.p. } 1-\epsilon \\ \sim \text{unif}(\calA), & \text{w.p. } \epsilon \end{cases} 
\end{equation}
where $Q_t(\bs_t,\ba) = \balpha_t^T\bk_t(\bs_t,\ba)$, and
\begin{equation*}
	N_t(\bs_t,\ba) = \sum_{\substack{(\bst_l,\bat_l) \in \mathcal{D}_t \\ \bat_l=\ba}} c_t(\bst_l,\bat_l) \exp\left( -\frac{\nn \bs_t-\bst_l \nn^2}{2\sigma_s^2} \right)
\end{equation*}
The dictionary of state-action pairs learned up to time $t$ is denoted as $\mathcal{D}_t$. Here $c_t(\bst_l,\bat_l)$ estimates the number of visits of state-action pair $(\bst_l,\bat_l)$ and is recursively computed as:
\begin{equation*}
	\bc_t = (1-\lambda) \bc_{t-1} + \bk_t(\bs_t,\ba_t)
\end{equation*}
if $\delta_t \leq \mu$, where $\bc_t$ is a $L$-dimensional vector of estimated counts. Here $\lambda\in (0,1)$ is a small positive parameter that controls the forgetting factor. If a new term is added, then $\bc_t = [\bc_{t-1}^T, 1]^T$. We refer to the guided exploration method (\ref{eq:akql}) as accelerated kernel Q-learning (AKQL). Kernel Q-learning (KQL) follows the same updated without the additional term $\frac{\max_{\ba} Q_t(\bs_t,\ba)}{N_t(\bs_t,\ba)}$ when exploiting.

\section{Budget-Constrained Dimensionality Reduction} \label{sec:dim_red}
The number of items in the nonparametric model (\ref{eq:kernelQ}) grows with time, increasing computational complexity and degrading generalization. It is critical to control the number of terms in $\mathcal{D}$ without losing important information. We propose an online sparsification technique based on a budget-constrained version of approximate linear dependence (ALD) analysis.

\subsection{Adding a new dictionary item}
ALD analysis \cite{Engel:2004, Xu:2007} is a method for sequentially building a dictionary of representative feature vectors $\phi(\bxt_i)=\phi(\bst_i,\bat_i)$. %When a state-action pair $\bx_t=(\bs_t,\ba_t)$ is observed, ALD provides a measure of how well $\phi(\bx_t)$ can be approximated as a linear combination of feature vectors already included in the dictionary.
Let the dictionary after testing $t-1$ samples be given by $\mathcal{D}_{t-1} = \{\bxt_j=(\bst_j,\bat_j)\}_{j=1}^{L_{t-1}}$. The new datum $\bx_t=(\bs_t,\ba_t)$ is included in the dictionary if $\delta_t = \min_{\bc} \parallel \sum_{j=1}^{L_{t-1}} c_j \phi(\bxt_j) - \phi(\bx_t) \parallel^2 > \mu$, where $\mu>0$ controls the approximation error. Defining the kernel matrix as $\bK_{t-1} = \{k(\bxt_i,\bxt_j)\}_{i,j=1}^{L_{t-1}}$ and the vector $\bk_{t-1}(\bx_t) = [k(\bxt_1,\bx_t),\dots,k(\bxt_{L_{t-1}},\bx_t)]^T$, we obtain $\delta_t = k(\bx_t,\bx_t) - \bk_{t-1}(\bx_t)^T \bK_{t-1}^{-1} \bk_{t-1}(\bx_t)$. If $\delta_t \leq \mu$, then $\mathcal{D}_t = \mathcal{D}_{t-1}$, and otherwise $\mathcal{D}_t = \mathcal{D}_{t-1} \cup \{\bx_t\}$. %The set of functions $\{\phi(\bx_l)\}_l$ constructed tends to be approximately linearly independent.

The ALD test may be recursively implemented by avoiding a direct inversion of the kernel matrix $\bK_{t-1}$. If a new state-action pair is added to the dictionary, the kernel matrix may be updated by adding a new row and column:
\begin{equation*}
	\bK_t = \left[ \begin{array}{c|c} \bK_{t-1} & \bk_{t-1}(\bx_t) \\ \hline \bk_{t-1}(\bx_t)^T & k_{tt} \end{array} \right]
\end{equation*}
The matrix inverse is then recursively updated as:
\begin{equation*}
	\bK_{t}^{-1} = \frac{1}{\delta_t} \left[
\begin{array}{c|c}
\delta_t \bK_{t-1}^{-1} + \bc_t \bc_t^T & -\bc_t \\ \hline
-\bc_t^T & 1
\end{array} \right]
\end{equation*}
with initial value $\bK_1^{-1} = [\frac{1}{k_{11}}]$. The ALD test then takes $O(L^2)$ time.
%ented using the recursive update, computational complexity to perform the ALD test is reduced to $O(L_{t-1}^2)$.

%This approach essentially sparsifies the kernel expansion in (\ref{eq:kernelQ}) by keeping relevant terms while ignoring terms that have been observed previously in the data stream and, more generally, can be well approximated by a linear combination of other terms in the dictionary.

\subsection{Removing a dictionary item}
Given a budget of $B$ state-action pairs due to processing and storage constraints, when $L=B+1$ one element must be removed from the dictionary. The first level of control is to remove any state-action pairs that have not been visited in a long time, i.e., if $\min_l N_t(\bst_l,\bat_l) \leq \epsilon$ for some $\epsilon \ll 1$, then $i_r = \arg \min_l N_t(\bst_l,\bat_l)$ is the item to be removed. If $\min_l N_t(\bst_l,\bat_l) > \epsilon$, then all state-action pairs are considered to have a high enough relative frequency and we choose to remove the feature vector that minimally distorts its local subspace.

Consider the removal of item $i$, $\phi_i = \phi(\bst_i,\bat_i)$. First we find the $k$ closest neighbors by partially sorting the vector $\bk(\bst_i,\bat_i)$ ($k$-th row of $\bK$) in a descending order, which takes $O(k L)$ time. This follows from the relation $\nn \phi(\bxt_i)-\phi(\bxt_l) \nn^2 = 2(1-k(\bxt_i,\bxt_l))$. Let us define the indexes of these neighbors as $\mathcal{J}_i$. Second we consider the subspace spanned by these vectors, $S_{\mathcal{J}_i} = \text{span}\left( \left\{ \phi(\bxt_l) \right\}_{l\in \mathcal{J}_i} \right)$. Removing $\phi_i$ distorts this subspace and we measure this distortion using kernel principal angles \cite{Wolf:2003}. Define the matrix $\bA_i$ with columns $\{\phi_l\}_{l\in \mathcal{J}_i}$ and $\bB_i$ with columns $\{\phi_l\}_{l \in \mathcal{J}_i \backslash \{i\}}$. The QR decompositions of the matrices $\bA_i=\bQ_{\bA_i}\bR_{\bA_i}, \bB_i=\bQ_{\bB_i}\bR_{\bB_i}$ can be used to compute the cosine principal angles as singular values of the matrix $\bQ_{\bA_i}^T\bQ_{\bB_i}$. Since $\bQ_{\bA_i}^T\bQ_{\bB_i} = \bR_{\bA_i}^{-T} \bA_i^T \bB_i \bR_{\bB_i}^{-1}$, and $\bA_i^T \bB_i = \bK_{\mathcal{J}_i,\mathcal{J}_i \backslash \{i\}}$, we may compute $\bR_{\bA_i}^{-1},\bR_{\bB_i}^{-1}$ using the efficient kernel Gram-Schmidt algorithm of \cite{Wolf:2003}. Finally we use the measure
\begin{equation*}
	d_i = \det([\bQ_{\bA_i}^T\bQ_{\bB_i}]_{1:k,1:k})^2 = \prod_j \cos(\theta_j)^2
\end{equation*}
as the subspace distortion measure due to removal of item $i$. This final step takes $O(k^3)$ time. The computational complexity for choosing the item to remove is $O(L(k L + k^3))$. In general $k \ll L$, so removal of a dictionary item takes $O(L^2)$ which matches the time complexity of ALD. The item to be removed is then $i_r = \arg \min_l d_l$.

When an item is removed, this corresponds to removing the corresponding column and row from the kernel matrix $\bK$. The inverse also needs to be updated. Assuming that the first row and column of $\bK$ are removed making up the smaller matrix $\bK_{\circ} = \bK_{2:L,2:L}$ of size $B\times B$ with inverse updated as:
\begin{equation*}
	\bK_{\circ}^{-1} = [\bK^{-1}]_{2:L,2:L} - [\bK^{-1}]_{2:L,1}[\bK^{-1}]_{2:L,1}^T/[\bK^{-1}]_{1,1}
\end{equation*}
Removing an arbitrary row and column from $\bK$ and similarly updating the inverse of $\bK_{\circ}$ is possible using the permutation technique of \cite{VanVaerenbergh:2010}.

\section{Numerical Experiments} \label{sec:simulation}

\subsection{Multi-channel Adversarial Data}

\subsubsection{Setup}
To compare the performance of the proposed algorithms we simulate a scenario in which a pair of SU cognitive radios attempt to operate in the presence of $K$ PU radio links. Each of the $K$ PU links consists of a pair of transceivers communicating over a channel with line-of-sight (LOS) propagation loss. The SU radios choose the channels to access over the next two slots by predicting the PU occupancy.

\begin{figure}[h]
	\includegraphics[width=0.50\textwidth]{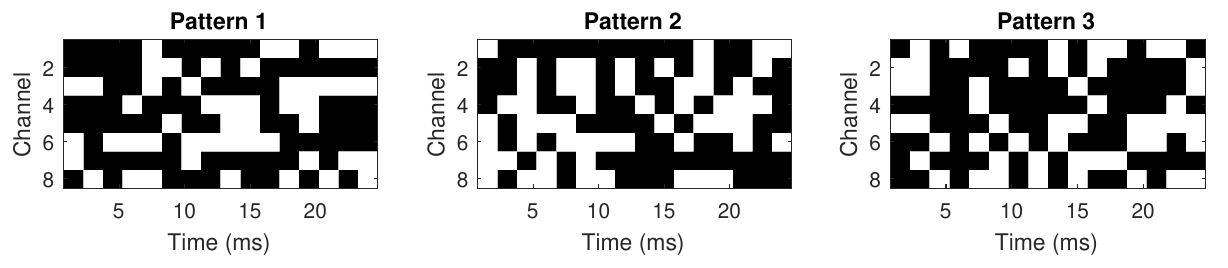}
	\caption{\small{Three communication patterns used by the adversarial PU network.}}
	\label{fig:patterns123}
\end{figure}

The SU and PU radios are assumed to operate at a rate of $R$ bits/sec\footnote{Specifically, each radio operates at the Nyquist rate of its respective channel using a $1/2$ rate code and antipodal signaling.} when no collisions occur. The values of each of the parameters chosen for the numerical experiments are summarized in Table \ref{tab:params}. %Throughout our experiments the radio links are configured in a particular randomly generated physical layout as shown in Figure (\ref{fig:netlayout}). 
\begin{table}
  \begin{tabular}{ c | c | c }
    \hline
    \textbf{Parameter} & \textbf{Value (units)} & \textbf{Description} \\ \hline \hline
    $K$ & $8$ (links or channels) & number of available channels\\ \hline
		$M$ & $20$ (slots) & number of time slots per state\\ \hline
    $W$ & $1.2$ (MHz) & bandwidth of each channel \\ \hline
    $R$ & $600$ (kbits/sec) & maximum data rate of each link\\ \hline
    $\Tslot$ & $1.5$ (ms) & slot duration \\ \hline
		$J$ & $2$ (slots) & channel access duration \\ \hline
    $P_{ptx}$ & $20$ (dBm) & PU network transmit power\\ \hline
    $(\sigma_s,\sigma_a)$ & (1.5, 0.2) & kernel parameters\\
    \hline
  \end{tabular}
  \caption{\small{Parameters used for coexistence with multi-channel adversarial radio.}}
	\label{tab:params}
\end{table}
The PU network is adversarial in the sense that it switches its pattern with probability equal to the access rate of the SU over the last packet. For time less than $1.5s$, patterns 1 and 2 shown in Fig. \ref{fig:patterns123} are switched, and for time greater than $1.5s$ patterns 2 and 3 are switched.

\subsubsection{Reward Design}
The reward encourages successful channel accesses and penalizes collisions with the PU, and is given by $r_t = +5R \sum_{l=MK-JK+1}^{MK} I_{\{\bs_{t+1}(l)<\tau, \ba_t(l)=1\}} - R \sum_{l=MK-JK+1}^{MK} I_{\{\bs_{t+1}(l)>\tau, \ba_t(l)=1\}}$
%\begin{align*}
	%r_t &= +5R \sum_{l=MK-JK+1}^{MK} I_{\{\bs_{t+1}(l)<\tau, \ba_t(l)=1\}} \\
		%&\quad - R \sum_{l=MK-JK+1}^{MK} I_{\{\bs_{t+1}(l)>\tau, \ba_t(l)=1\}}
%\end{align*}
where $R$ is the max data rate of each link shown in Table \ref{tab:params} and $\tau$ is the occupancy threshold. Here successful channel accesses are weighted $\times 5$ more than collisions.

%\subsubsection{State/Action Considerations}
%The state $\bs_t$ consists of power measurements obtained over the last $M$ time slots, making up a vector of size $MK$. The next state $\bs_{t+1}$ is obtained by shifting over $J$ time slots, as the action duration is made up of $J$ slots. Prior to being processed by any of the learning algorithms, the observed power measurements are normalized by mapping them to the unit interval.

\subsubsection{Results}
\begin{figure}[h]
\centering
	\includegraphics[width=0.50\textwidth]{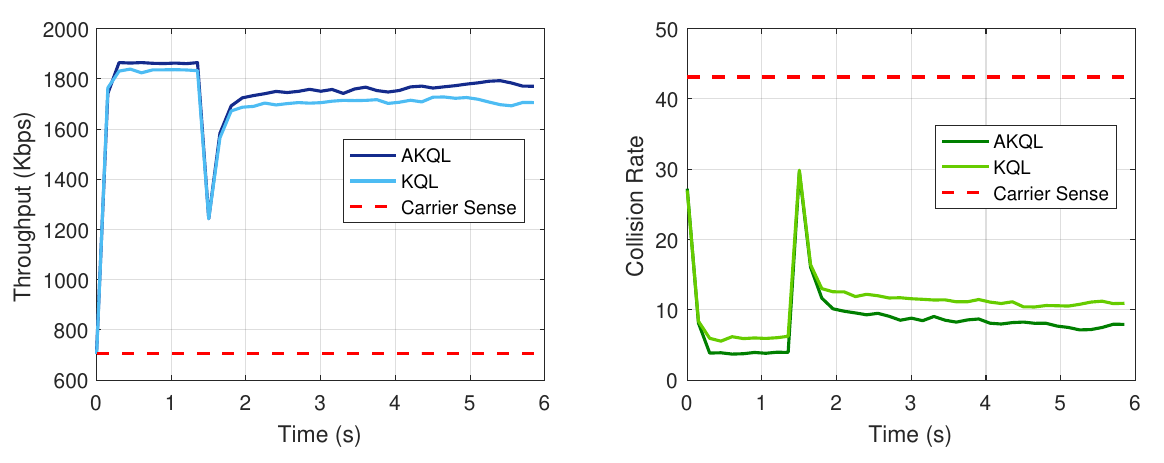}
	\caption{\small{Throughput and collision rate for accelerated kernel Q-learning (AKQL), kernel Q-learning (KQL), and carrier sense.}}
	\label{fig:akql_kql_offline}
\end{figure}

The SU throughput and collision rate as a function of time are shown in Fig. \ref{fig:akql_kql_offline}, averaged over $30$ trials. We observe that the RL techniques significantly outperform carrier-sense based methods in terms of throughput and collision rate during a short period of time. Furthermore, due to the abrupt change in PU network communication patterns at $1.5s$, the RL techniques both recover fast by forgetting irrelevant state-action pairs and learning new ones given a budget of $B=1000$ state-action pairs. The guided exploration of AKQL achieves a higher throughput and lower collision rate than KQL consistently. The maximum achievable throughput here is $1950$Kbps, implying a performance above $92\%$ optimal throughput achieved at a collision rate below $5\%$.

\subsection{Single-channel Passive Data}
%We illustrate the performance of our KQL approach in a single frequency channel environment with realistic network traffic.
\subsubsection{Setup}  We simulate the performance of KQL when operating in the presence of a pair of PU radios that employ a carrier-sense multiple-access with collision avoidance (CSMA-CA) protocol, which is widely used in wireless local area networks (WLANs) \cite{80211p11}, and wireless personal area networks (WPANs) \cite{802154}.

%The SU receiver runs the KQL algorithm with $\epsilon$-greedy action selection as detailed in Section \ref{sec:KQL_GE}. 
We assume a free-space LOS propagation environment where the PUs are observed by the SU receiver at signal-to-noise ratios (SNRs) specified in Table \ref{tab:params_sc}. To illustrate the benefit of using KQL in the presence of CSMA-CA networks, we compare its performance to that of a SU link employing a simple carrier sensing (CS) method known as $p$-persistent CSMA, without collision avoidance \cite{pcsma}.

\subsubsection{Reward Design}The CSMA-CA protocol uses carrier sensing and an exponential backoff method to adapt to co-channel network traffic. Because of this, the PU network tends to yield use of the channel to interfering users. This passive behavior can lead the SU to learn a policy that dominates the channel and allows little or no PU accesses to occur. To prevent the SU from learning such an unfair policy, the reward function is designed as a function of $\ba_t$, $\bs_{t+1}$, and the current number of consecutive channel access actions by the SU, denoted $n_{t}$. 

When the SU chooses to access the channel at time $t$, the reward function is
\begin{equation*}
	r_t = \begin{cases} +(1-\nu)c f(n_{t};N_p), & \text{if final $\Sca$ slots in $\bs_{t+1}$ are idle}  \\ -2c, & \text{else} \end{cases}
\end{equation*}\label{eq:reward_sc_1}
where $c>0$ and $0 < \nu < 1$ are positive constants, $n_t, N_p \in \NNz$, and $f(n_t;N_p) = \tanh\left(N_{p} - n_{t}\right)$, causing the reward to become negative when $n_t > N_{p}$. 

When the SU chooses to remaining idle at time $t$, the reward function is
\begin{equation*}
	r_t = \begin{cases} +\nu c, & \text{if any of final $\Sca$ slots in state $\bs_{t+1}$ are busy}  \\ +\zeta, & \text{else} \end{cases}
\end{equation*}\label{eq:reward_sc_2}
where $\zeta>0$ is a small positive constant. We remark that $\nu$ is used to help determine the relative priority of SU vs PU network traffic, whereas $\zeta$ gives the SU network incentive to let the channel remain idle for short periods of time, which allows the CSMA-CA backoff time to elapse.

\begin{table}
  \begin{tabular}{ c | c | c }
    \hline
    \textbf{Parameter} & \textbf{Value (units)} & \textbf{Description} \\ \hline \hline
    $K$ & $1$ (links or channels) & number of available channels\\ \hline
    $M$ & $25$ (slots) & number of time slots per state\\ \hline
    $W$ & $15$ (MHz) & bandwidth of each channel \\ \hline
    $R_{g}$ & $12.5$ (Mbits/sec) & max. data rate of PU network\\ \hline
    $R_{bcs}$ & $23.4$ (Mbits/sec) & max. data rate of carrier sense radio\\ \hline
    %$R_{brl}$ & $x$ (kbits/sec) & max. data rate of RL radio link\\ \hline
    $\Tslot$ & $20$ ($\mu$s) & slot duration\\ \hline
    $\Sca$ & $10$ (slots) & channel access duration\\ \hline
    $\bm{\rho}_g$ & $\{18,20\}$ (dB) & observed SNR of PU radios\\ \hline
    $\bm{\rho}_b$ & $19$ (dB) & observed SNR of SU transmitter\\ \hline
    $(\nu,c,\zeta)$ & $(0.4,1,0.01)$ & Reward parameters\\ \hline
    $(\sigma_s,\sigma_a)$ & $(0.225,P_{stx}^{1/2})$ & kernel parameters\\
    \hline
  \end{tabular}
  \caption{\small{Parameters used for coexistence with single-channel CSMA-CA network.}}
  \label{tab:params_sc}
\end{table}

Note that varying $N_{p}$ controls the value corresponding to different numbers of consecutive channel accesses by the SU\footnote{$n_t$ is reset to $0$ whenever the SU selects the idle action.}, and can thus be used to control a SU-PU performance tradeoff. In our comparison, a similar SU-PU performance tradeoff is achieved by varying the carrier sense parameter $p \in [0,1]$.

%\subsubsection{State/Action Considerations}
%We include $n_{t}$ in the state as an auxiliary term, along with a sliding temporal window of $M$ power measurements observed at the SU receiver. Note that the length $M$ window of power measurements shifts forward in time by $\Sca$ time slots with each successive state due to the duration of each action. As in the multichannel case, the power measurements have been normalized to lie in the unit interval.

\subsubsection{Results}
\begin{figure}[h]
	\includegraphics[width=0.50\textwidth]{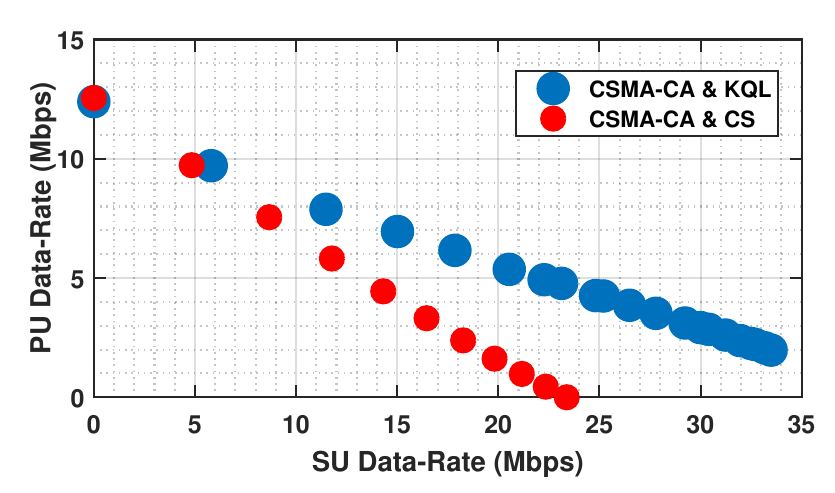}
	\caption{\small{Data rate tradeoff for a two-user PU network employing CSMA-CA under two scenarios where the SU employs either: (blue) KQL, and (red) carrier sense.}}
	\label{fig:csmaca_tradeoff1}
\end{figure}

\begin{figure}[h]
	\includegraphics[width=0.50\textwidth]{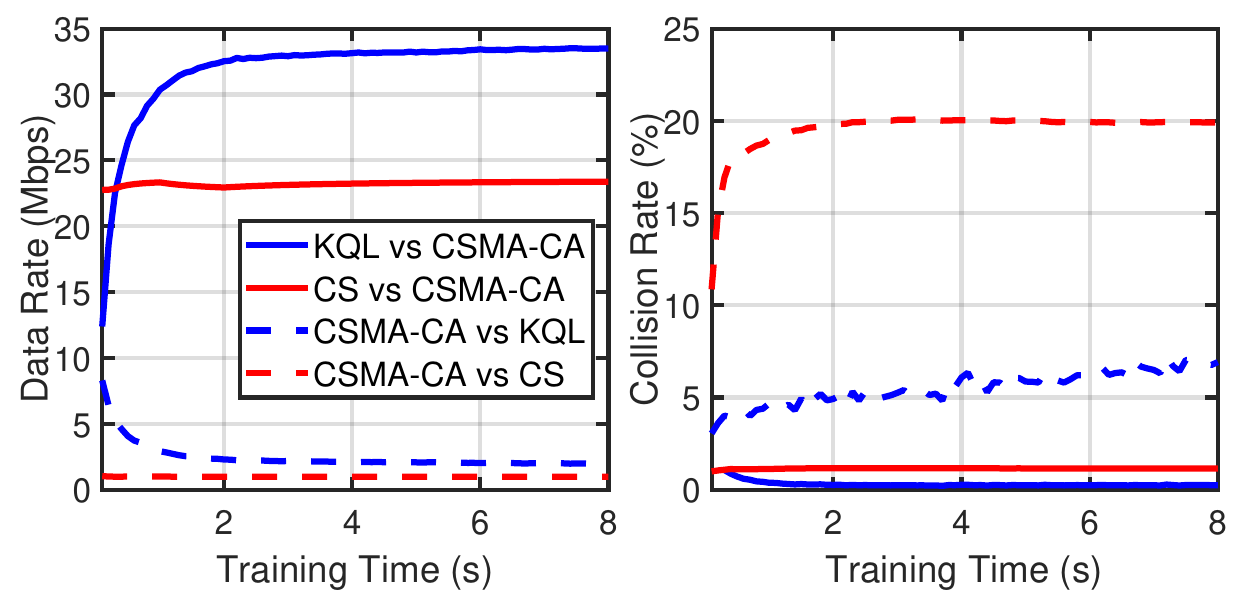}
	\caption{\small{Data and collision rate as a function of time spent training the RL algorithm for a two-user PU network employing CSMA-CA under two scenarios where the SU employs either: (blue) KQL, and (red) carrier sense. Solid lines correspond to SU data and collision rates, and dashed lines to PU data and collision rates.}}
	\label{fig:csmaca_train_datarate1}
\end{figure}
The SU-PU data rate tradeoff is shown in Fig. \ref{fig:csmaca_tradeoff1}.  As the SU data rate increases, the advantage of using KQL over a simple carrier sense approach becomes significant. For example, we see that for a desired SU data rate of $20$Mbps, KQL enables a greater than $3$x increase in PU data rate. As the PU data rate increases, KQL and carrier sense perform similarly. Figure \ref{fig:csmaca_train_datarate1} illustrates the data and collision rates for a particular SU-PU set along the tradeoff. Note that collision rates of the SU and PU are computed as $\frac{N_{col}}{N_{SU}}$ and $\frac{N_{col}}{N_{PU}}$, respectively, where $N_{SU}$ and $N_{PU}$ are the number of packet transmission attempts by the PU and SU, and $N_{col}$ is the number of collisions. The advantage of using KQL in the SU radio when high SU data rates (e.g. $>30$Mbps) are desired is apparent. Note the collision rates of the SU and PU are both lower when KQL is used by the SU. We remark that we did not observe the state action dictionary size to reach the budget constraint of $B=1000$ in this single channel scenario. %Study of the multichannel CSMA-CA scenario, where the budget constraint would become applicable, is differed to future work.%This is to be expected, since both SU approaches allow the PU to utilize an increasing amount of the available channel uses.
\vspace{-10pt}
\section{Conclusion}
We used kernel-based reinforcement learning techniques for predicting and accessing idle spectrum. An acceleration heuristic and a novel budget-constrained dimensionality reduction method was proposed to control the model complexity. Numerical experiments show improvements in throughput and collision rate over carrier-sense systems for coexistence with passive and adversarial radio networks.

%\vfill\pagebreak

% References should be produced using the bibtex program from suitable
% BiBTeX files (here: strings, refs, manuals). The IEEEbib.bst bibliography
% style file from IEEE produces unsorted bibliography list.
% -------------------------------------------------------------------------
\bibliographystyle{IEEEbib}
\bibliography{refs}

\end{document}